\documentclass{pasj00}
\draft

\begin{document}
\SetRunningHead{Morii et al.}{4U 0142+61}
\Received{2008/06/16}
\Accepted{2008/09/09}

\title{Search for Near-Infrared Pulsation of the Anomalous X-ray Pulsar 4U~0142+61}


\author{
	Mikio \textsc{Morii}\altaffilmark{1},
	Naoto  \textsc{Kobayashi}\altaffilmark{2},
	Nobuyuki \textsc{Kawai}\altaffilmark{3},
	Hiroshi \textsc{Terada}\altaffilmark{4}, \\
        Yasuyuki T. \textsc{Tanaka}\altaffilmark{5},
	Shunji \textsc{Kitamoto}\altaffilmark{1},
	and
	Noriaki \textsc{Shibazaki}\altaffilmark{1}
	}

 \altaffiltext{1}{Department of Physics, Rikkyo University, Nishi-ikebukuro 3-34-1, \\
Toshima-ku, Tokyo 171-8501, Japan}
 \email{mmorii@rikkyo.ac.jp}
 \altaffiltext{2}{Institute of Astronomy, Graduate School of Science, University of Tokyo,
	2-21-1 Osawa, \\
Mitaka, Tokyo 181-0015, Japan}
 \altaffiltext{3}{Department of Physics, Tokyo Institute of Technology,
   Ookayama 2-12-1, Meguro-ku, \\
 Tokyo 152-8551, Japan}
 \altaffiltext{4}{Department of Earth and Planetary Science, University of Tokyo,
   7-3-1 Hongo, Bunkyo-ku, \\
Tokyo 113-0033, Japan}
 \altaffiltext{5}{Subaru Telescope, National Astronomical Observatory of Japan,
 650 North A'ohoku Place, \\
Hilo, HI 96720, USA}

%


\maketitle

\begin{abstract}
We have searched for pulsation of the anomalous X-ray pulsar (AXP) 4U 0142+61
in the $K^\prime$ band ($\lambda_{\rm eff} = 2.11$ $\mu$m)
using the fast-readout mode of IRCS at the Subaru 8.2-m telescope.
We found no significant signal at the pulse frequency
expected by the precise ephemeris obtained by the X-ray monitoring observation with RXTE.
Nonetheless, we obtained a best upper limit of 17\% (90\% C.L.)
for the root-mean-square pulse fraction in the $K^\prime$ band.
Combined with $i^\prime$ band pulsation \citep{Dhillon+_2005},
the slope of the pulsed component ($F_\nu \propto \nu^\alpha$)
was constrained to $\alpha > -0.87$ (90\% C.L.) for an interstellar extinction of $A_{V} = 3.5$.
\end{abstract}

\section{Introduction}

Anomalous X-ray pulsars (AXPs; see \citet{Woods_Thompson_2006}
for a recent review) have been recognized as being 
enigmatic X-ray pulsars, for which none of characteristics
of binary companions have been detected \citep{Mereghetti_Stella_1995}.
Therefore, the observed X-ray luminosity of AXPs
($L_{\rm X} \sim 10^{34} - 10^{36}$ erg s$^{-1}$)
cannot be explained by the accretion of matter from companions,
as is the case for binary X-ray pulsars.
The spin periods of AXPs ($P$) are concentrated in a narrow range
(2$-$12 s \footnote{Recently, the period range was extended
down to 2 s by the discovery of the radio-emitting AXP 1E 1547.0$-$5408 \citep{Camilo+_2007}.})
and the spin-down rates ($\dot{P}$) are distributed in the range
$5 \times 10^{-13}$ -- $1 \times 10^{-10}$ s s$^{-1}$.
The rate of the spin-down energy loss of neutron stars
($\dot{E} = - 4 \pi^2 I \dot{P}/P^3 \sim 10^{32.6}$ erg s$^{-1}$,
where $I \simeq 10^{45}$ g cm$^2$ is 
the moment of inertia of a neutron star)
is too small to maintain the X-ray luminosity,
suggesting that they are not rotation-powered pulsars.

Two competing models have been proposed for the energy source:
accretion from a fossil disk
and decay of ultra-strong magnetic fields.
In the former scenario, the fossil disks are thought to be produced
through a fall-back of matter after a supernova explosion
\citep{Chatterjee_Hernquist_Narayan_2000, Chatterjee_Hernquist_2000, Alpar_2001}
or as a remnant of the common-envelope phase in the evolution of a binary
\citep{van_Paradijs_Taam_van_den_Heuvel_1995, Ghosh_Angelini_White_1997}.
The advantage of the fossil disk model is that
it naturally explains the narrow range of the concentration of the spin periods
through the propeller phase in the spin evolution
\citep{Chatterjee_Hernquist_Narayan_2000, Chatterjee_Hernquist_2000, Alpar_2001}.
In the latter scenario AXPs are magnetars,
which are strongly magnetized neutron stars
with emissions powered by the dissipation of magnetic energy
\citep{Thompson_Duncan_1995, Thompson_Duncan_1996}.
Assuming that their spins are braked by magnetic dipole radiation,
a strong magnetic field of $10^{14}$ -- $10^{15}$ G at their surface 
is implied.
AXPs are similar to soft gamma repeaters (SGRs),
which are thought to be magnetars (e.g. see \cite{Thompson_Duncan_1995}).
The observation of SGR-like bursts in AXPs suggests the unification of AXPs and SGRs
by the magnetar scenario \citep{Gavriil_Kaspi_Woods_2002}.

4U 0142+61 is a typical object among AXPs.
It is the nearest and least absorbed
object among AXPs and SGRs and is hence
the most intensively observed of these objects
in multi-wavelength studies.
The optical counterpart of this object was
discovered by \citet{Hulleman_van_Kerkwijk_Kulkarni_2000}.
\citet{Kern_Martin_2002} discovered the optical pulsation.
The pulsed fraction of $PF_{\rm opt} = 27^{+8}_{-6}$\%
was found to be larger than that of soft X-rays,
suggesting that the optical pulsation cannot be explained by
thermal reprocessing of a disk illuminated by X-rays.
\citet{Dhillon+_2005} also measured
the optical pulsation in a $i^\prime$ band
and obtained a consistent pulsed fraction of $PF_{i^\prime} = 29 \pm 8$\%.

We showed that 
the near-infrared/optical spectrum consists of two components,
near-infrared excess and optical pulsating components,
by first near-simultaneous multi-band photometry
\citep{Morii+_2005, Tanaka+_2008}.
\citet{Wang_Chakrabarty_Kaplan_2006} discovered
the mid-infrared emission with Spitzer, and
showed that the near-infrared excess component extends to
the mid-infrared region.
They proposed that the infrared excess component can be modeled
by a dust disk around a neutron star,
where the spectrum is similar to that of the dust disk of a proto-planetary system.
Their model is different from the original fossil disk model,
in that the disk does not power the X-ray emission.
In their model, the flux of the $K$ band is mainly composed of the dust
disk component. Therefore, the pulsed flux of the $K$ band is expected to be small.
\citet{Durant_van_Kerkwijk_2006b} collected data spanning seven years
and showed that the near-infrared/optical flux and the spectral shape are variable.
The correlations between the X-ray total/pulsed fluxes and the near-infrared/optical fluxes
were uncertain, although these correlations are expected for a fossil disk.
Therefore, they suggested that the near-infrared flux is composed of components
other than the disk component.
As described above, the origin of the infrared emission is not yet clear.

Some authors have proposed theoretical models for the optical/infrared emission of AXPs,
based on the magnetar scenario or fossil disk scenario.
In the former scenario,
\citet{Eichler_Gedalin_Lyubarsky_2002, Lu_Zhang_2004} applied
mechanisms of coherent radio emission of rotation-powered pulsars to a magnetar,
where the emission frequency is boosted to the infrared, optical, or even UV
region by the strong magnetic field of the magnetar.
\citet{Heyl_Hernquist_2005} proposed an exotic quantum electrodynamics model
for the broad-band emission of magnetars.
\citet{Ertan_Cheng_2004} applied the outer gap model
of rotation-powered gamma-ray pulsars on magnetars.
In the latter scenario,
\citet{Perna_Hernquist_Narayan_2000, Perna_Hernquist_2000, Hulleman_van_Kerkwijk_Kulkarni_2000}
discussed the optical/infrared spectrum generated by an irradiated fall-back disk
and found that the sizes of the inner and outer radii of the disk become implausible.
In addition, hybrid models of both scenarios have also been proposed.
\citet{Ertan_Cheng_2004} showed that the pulsed optical emission can be
explained by the disk-star dynamo gap model, where
the fossil disk plays a key role in the emission mechanism.
\citet{Ertan_Caliskan_2006, Ertan+_2007} showed
that an irradiated and viscously heated disk model can account
for the observed infrared/optical spectrum
of both pulsed and un-pulsed components,
by magnetospheric optical emission \citep{Ertan_Cheng_2004}
and disk emission, respectively.

In any of the above models,
the origin of the emission can be clearly determined by the pulsation.
The emission from the co-rotating magnetosphere must be pulsed,
while that from the disk is unlikely to be pulsed.
In this paper, we report a search for the pulsation
of the AXP 4U 0142+61 in the $K^\prime$ band,
using the fast-readout mode of IRCS mounted on the Subaru 8.2-m telescope.
This is the first search for the pulsation in the infrared band.

In the following, the dereddened spectrum of 4U 0142+61 is obtained
using the interstellar extinction of $A_V = 3.5 \pm 0.4$,
which was determined by using red clump stars
\citep{Durant_van_Kerkwijk_2006a}.

\section{Observation} \label{section: observation}

We observed the AXP 4U 0142+61 in the $K^\prime$ band 
(MKO-NIR photometric system; \cite{Simons_Tokunaga_2002}) on 2004 July 31 (UT),
using IRCS (Infrared Camera and Spectrograph)\citep{Kobayashi+_2000, Tokunaga+_1998}
mounted on the Cassegrain focus of the Subaru Telescope \citep{Iye+_2004}
on top of Mauna Kea, Hawaii.
The weather condition was photometric with an excellent seeing
($\sim0.''4$ at the $K^\prime$ band) through the observing.
In the MKO-NIR photometric system,
the effective wavelength and band width of the $K^\prime$ band
are $\lambda_{\rm eff} = 2.11$ $\mu$m and $\Delta \lambda = 0.34$ $\mu$m (FWHM),
respectively \citep{Tokunaga_Simons_Vacca_2002, Tokunaga_Vacca_2005}.
To search for pulsation of the AXP, we used IRCS in the {\it movie mode}.

In this mode, we successively took 0.84-s exposure frames,
corresponding to about 1/10 of the pulse period of the AXP (8.7 s).
To reduce the overhead time for array readout and data recording,
we restricted the readout region to
the central sub-array of $256 \times 256$ pixels,
which corresponds to 1/8 of the $1024 \times 1024$ pixels of the full image.
We used a pixel scale of 58 mas per pixel, and hence the $256 \times 256$-pixel
region corresponded to a field of view (FoV) of $14^{\prime\prime}.8 \times 14^{\prime\prime}.8$.

We observed the AXP by two-point dithering.
We arranged the FoV of IRCS so that
at least one of two bright stars near the AXP
(109 and 115 listed in \cite{Hulleman_van_Kerkwijk_Kulkarni_2004})
were included in the FoVs of both observation directions.
We call one of the two directions, for which star 109 and the AXP were included
in the FoV, the ``A direction,'' and
we call the other, for which stars 115 and 109 and the AXP were included in the FoV,
the ``B direction.''
These stars were bright enough to be seen in all frames.
We successively took 300 frames for each direction, saved into a FITS file
of cube structure, called a ``data-cube'', and recorded
the instantaneous time of each frame.
The observation directions were changed in the order A, A, B, B, A, A, \dots.
Good quality data taken in our observation are summarized in table \ref{tab: obs summary}.



\begin{table}
  \caption{Summary of our observation}\label{tab: obs summary}
  \begin{center}
    \begin{tabular}{llll}
\hline
Date (UT)   &  2004-07-31 \\
Start time (UT) & 12:57:14 \\
End time (UT)   & 15:21:07 \\
Duration    &  8633 s  \\
Filter      & $K^\prime$ ($\lambda_{\rm eff}: 2.11$ $\mu$m) \\
NDR         &   16        \\
Co-adds     &   1         \\
Exposure    &  0.84  s    \\
\# of frames & $300 \times 22 = 6600$ \\
Net exposure &  5.5 ks  \\
Seeing (FWHM) & 0.$^{\prime\prime}$37 \\ \hline
    \end{tabular}
  \end{center}
\end{table}

\section{Analysis}

\subsection{Search for Pulsation} \label{subsection: search pulse}

The precise pulse frequency of 4U 0142+61 has been measured with
RXTE over a period of ten years \citep{Dib_Kaspi_Gavriil_2007}.
The period of our Subaru observation is included in this interval.
The pulse frequency of this AXP at the start of our Subaru observation
was 0.1150934471(15) Hz, where the number in parentheses is the 1$\sigma$ uncertainty.
Here, we used the ``Postgap ephemeris spanning cycle 5--10'' in
table 2 of \citet{Dib_Kaspi_Gavriil_2007}.

We searched for pulsation in our Subaru data near this pulse frequency
by the epoch folding method
from 0.113601 Hz to 0.116601 Hz in steps of 0.00001 Hz.
The range and step were sufficiently wide and fine, respectively,
because the Fourier resolution of the pulse frequency
%
%
was $\Delta \nu = 1/2T = 5.8 \times 10^{-5}$ Hz,
where $T$ is the observation span (table \ref{tab: obs summary}).
%
Since the pulse period of the AXP is long (8.7 s)
and the duration of the observation was short (2.4 hours),
the only significant effect of the Doppler shift due to the motion of observer
was that due to the orbital motion of the earth,
estimated to be
$\Delta \nu = + 7.6 \times 10^{-6}$ Hz.
This is small in comparison with the Fourier resolution.
Thus, barycentric correction of the arrival times of photons was not necessary
in our observation.

We analysed the data with the IRAF
software packages.
We decomposed the data-cubes of the A and B directions
into normal FITS images, ${\rm A}_i$ and ${\rm B}_i$ ($i = 1, 2, \cdots, 300$).
The first image of each data-cube was excluded, because 
its quality was poor owing to bias instability.
We produced ${\rm A}_i - {\rm B}_i$ images ($i = 2, 3, \cdots, 300$),
by subtracting a frame of the B direction from a frame of the A direction.
By this process, the sky background and the dark level were canceled.
The pair of data-cubes were selected with
as small an interval between the directions as possible
to cancel the sky background properly.
Subtraction between frames of the same index $i$ was
necessary to cancel the sequencing variation of the bias.
${\rm B}_i - {\rm A}_i$ images were also produced in the same way.
%
%
Flat fielding was performed for the ${\rm A}_i - {\rm B}_i$ and ${\rm B}_i - {\rm A}_i$ images
using a mean-combined flat frame taken by irradiation of a calibration lamp.

We sorted the flat-fielded images into eight phases
of a tentative pulse period of the AXP,
according to the instantaneous times at their exposures.
For each phase, we shifted and combined the flat-fielded images
by masking the bad pixels.
The map of the bad pixels was made from the dark and flat frames,
for which we selected pixels with extraordinary ADU counts.
The number of pixels for the image-shift was calculated
by measuring the center position of the point spread function of
the two bright stars (109 and 115; see section \ref{section: observation}).



\begin{figure}
  \begin{center}
    \FigureFile(60mm,60mm){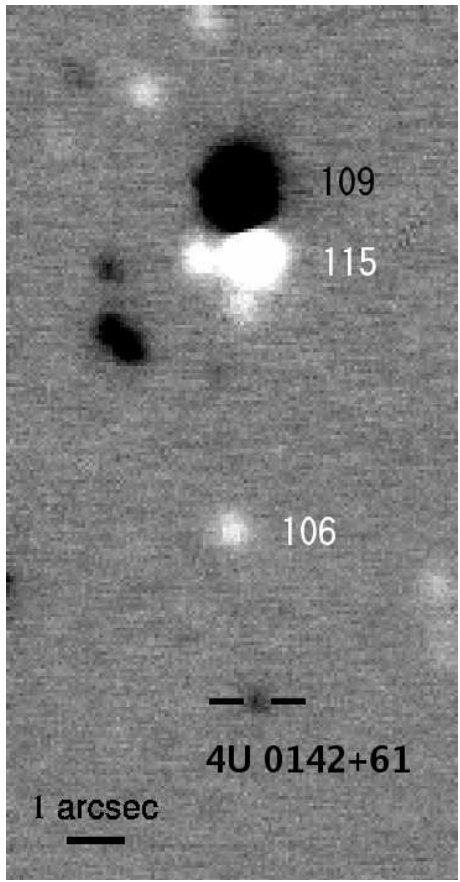}
  \end{center}
  \caption{$K^\prime$ image of the field around 4U 0142+61, generated by combining all the
	images of the IRCS movie mode. A 1$''$ scale is shown in the
	bottom-left corner of the image.
	4U 0142+61 is indicated by two horizontal bars with a spacing of 8 pixels
	($0.''46$).
	The brightest star in the upper region is star 109 (see section \ref{section: observation}).
	The dimples of inverted color are side-effects of the generation of
	the ${\rm A}_i - {\rm B}_i$ and ${\rm B}_i - {\rm A}_i$ images.
	For example, dimples corresponding to stars 115 and 106 were
	appeared by processes ${\rm A}_i - {\rm B}_i$ and ${\rm B}_i - {\rm A}_i$,
	respectively.}
	\label{fig: 4U+mark_08030300_crop3.eps}
\end{figure}

\begin{figure}
  \begin{center}
    \FigureFile(80mm,80mm){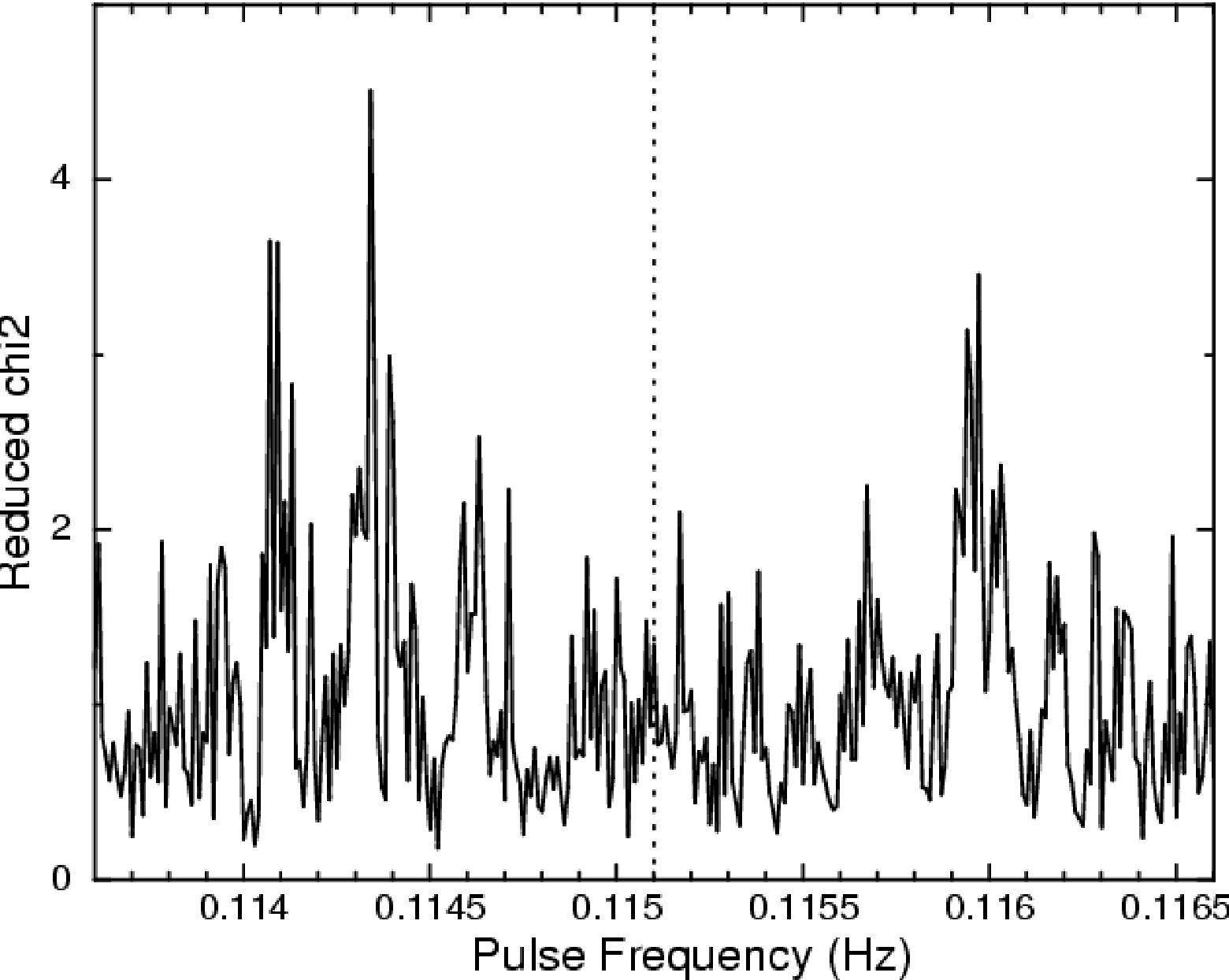}
  \end{center}
  \caption{Result of pulse search for an aperture radius of 4 pixels (0.$^{\prime\prime}$23).
        The vertical dotted line denotes the expected pulse frequency (see text).}
	\label{fig: periodgram_4U_pulse_v7.0_paper.ps}
\end{figure}

We measured the flux (ADU) of the AXP of each phase
using the standard IRAF photometry package (PHOT).
For the aperture, we set
the radii from 1 pixel (0$^{\prime\prime}$.058)
to 15 pixels (0$^{\prime\prime}$.87) in steps of 1 pixel.
For the background region, we set a concentric annulus
with radii of the inner and outer boundaries of
15 and 25 pixels (1$^{\prime\prime}$.45).
%
%
%
The systematic error of the flux was estimated by photometry for 11 areas of blank sky
for the same aperture and background annulus as the area of the AXP.
In the following, we applied an aperture radius of 4 pixels (0.$^{\prime\prime}$23).
For an aperture radius of around 4 pixels, the signal-to-noise ratio
of the flux within the aperture to
the systematic error obtained as above became maximum
for a phase-averaged image (figure \ref{fig: 4U+mark_08030300_crop3.eps}).
We produced a pulse profile with eight phase bins, and calculated
the reduced $\chi^2$ values for the profile
by the best fit constant model.
The reduced $\chi^2$ value represents the signal of the pulsation of the AXP.
Figure \ref{fig: periodgram_4U_pulse_v7.0_paper.ps} shows
the signal versus the pulse frequency.
We could not detect any significant signal of pulsation of the AXP at
the expected pulse frequency in any aperture radius setting.

In addition, we tried a similar pulse search
by relative photometry with reference to the flux of
stars 109 or 106 listed in \citet{Hulleman_van_Kerkwijk_Kulkarni_2004}
(figure \ref{fig: 4U+mark_08030300_crop3.eps}).
For star 109, the values in a plot
corresponding to figure \ref{fig: periodgram_4U_pulse_v7.0_paper.ps}
was larger than that obtained by the simple photometry above.
It is thought that there is an adverse influence of
the dimple of star 115 close to star 109.
In contrast, for star 106 the reduced $\chi^2$ values were slightly smaller,
indicating a slight improvement of the photometry.
Since this improvement was small,
we decided to apply simple photometry
rather than relative photometry with reference to star 106.


%
%


\subsection{Upper Limit for Pulsation}

The infrared pulse profile of 4U 0142+61 is unknown.
Then, at first we calculated an upper limit of the pulse fraction
by the root-mean-square pulse fraction \citep{van_der_Klis_1989}:
\[
PF_{\rm rms} = \frac{ \sqrt{ \frac{1}{N} \sum_i (R_i - \bar{R})^2 } }{\bar{R}},
\]
where $R_i$ ($i = 1, 2, \dots, N; N = 8$) and $\bar{R}$ are 
the flux of 4U 0142+61 of the $i$-th phase bin and
the mean of $R_i$, respectively.
We produced a histogram of $PF_{\rm rms}$ of a pulse profile
for the searched pulse frequencies, and hence the $90$\% C.L. upper limit
was obtained from the distribution of the histogram.
We followed this calculation for each aperture radius,
and the least upper limit was obtained at
an aperture radius of 5 pixels (0.$^{\prime\prime}29$).
The 90\% C.L. upper limit of the pulse fraction was 17\%.
%
%

Assuming the pulse profile, the tighter upper limit can be obtained.
Because of the non-detection of the infrared pulsation,
an assumption of a sinusoidal shape is generally acceptable.
For 4U 0142+61, both single-peaked and double-peaked shapes are plausible.
For example, the optical pulse profile shown in \citet{Kern_Martin_2002}
is a double-peaked shape, while that shown in \citet{Dhillon+_2005}
looks like a single-peaked shape.
Then, we fitted the pulse profile by a sine curve
with three free parameters ($PF_{\rm sin}$, $B$ and $\delta$),
\[
F = B \times \left[ PF_{\rm sin} \sin (2 \pi \phi + \delta) + 1 \right] 
\]
or
\[
F = B \times \left[ PF_{\rm sin} \sin (4 \pi \phi + \delta) + 1 \right] 
\]
with the $\chi^2$ fitting method
for each pulse frequency and each aperture radius.
Here, $\phi$ is the pulse phase between 0 and 1.
We followed the same method as that described in $PF_{\rm rms}$
for the calculation of the $90$\% C.L. upper limit of the $PF_{\rm sin}$.
The least 90\% C.L. upper limit of the pulse fraction was both 16\%
at aperture radii of 5 pixels and 3 pixels (0.$^{\prime\prime}$17), respectively.
%
%

Eventually, the resultant upper limit of the pulse fraction is 17\% (90\%C.L.)
by selecting a conservative value.

\subsection{Phase-averaged flux} \label{subsection: Phase-averaged flux}

The phase-averaged flux of 4U 0142+61 was
obtained by relative photometry of an image made by combining all the images of
the movie mode (figure \ref{fig: 4U+mark_08030300_crop3.eps}).
The magnitude of the AXP was measured with reference to
the magnitude of star 106 listed in \citet{Hulleman_van_Kerkwijk_Kulkarni_2004}.
The aperture radius and the radii of the inner and outer boundaries of
the concentric annulus for the background were set to 4, 15 and 25 pixels, respectively.
The systematic error for the flux was estimated 
as described in subsection \ref{subsection: search pulse}.
The resultant magnitude of the AXP was $K^\prime = 19.73 \pm 0.09 \pm 0.03$,
where the last error is the zero-point uncertainty inherited from
the catalog \citep{Hulleman_van_Kerkwijk_Kulkarni_2004}.

\section{Discussion}

\begin{figure}
  \begin{center}
    \FigureFile(80mm,80mm){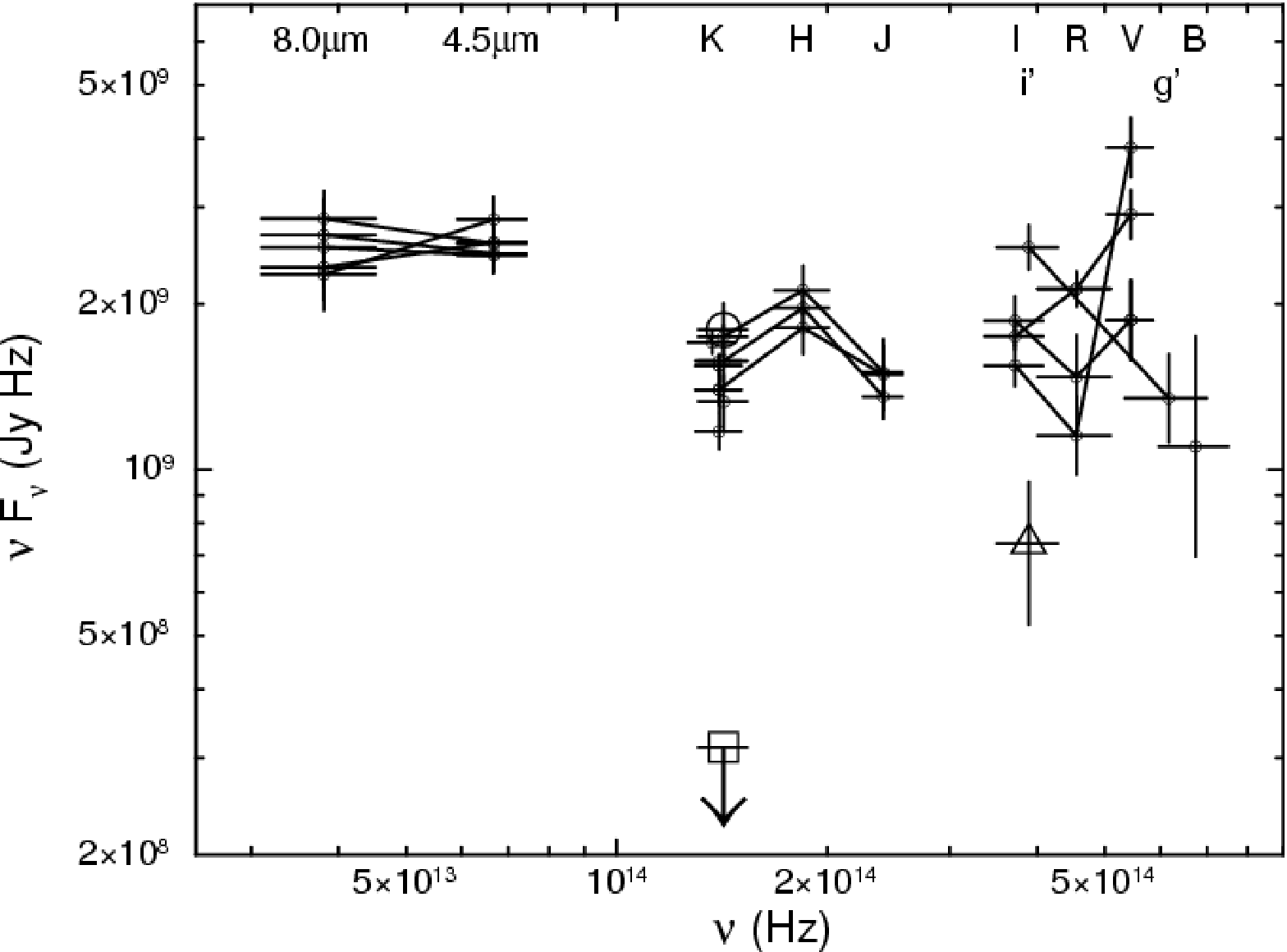}
  \end{center}
  \caption{Energy spectrum ($\nu F_\nu$) of 4U 0142+61.
The horizontal and vertical axes are shown in units of Hz and Jy$\cdot$Hz, respectively.
Phase-averaged fluxes derived from \citet{Hulleman_van_Kerkwijk_Kulkarni_2000,
Hulleman_van_Kerkwijk_Kulkarni_2004,
Morii+_2005,
Durant_van_Kerkwijk_2006b,
Wang_Chakrabarty_Kaplan_2006,
Wang_Kaspi_2008,
Tanaka+_2008}
are marked as small circles.
The flux obtained by this work (see subsection \ref{subsection: Phase-averaged flux})
is marked as a large circle.
The pulsed flux in the $i^\prime$ band is derived
from \citet{Dhillon+_2005} and marked as a large triangle.
The $90$\% C.L. upper limit for the pulsed flux in the $K^\prime$ band
obtained in this work ($PF_{\rm rms}$)
is marked as a large square and downward arrow.
The pulsed fluxes are calculated as the product of
the phase-averaged flux and the pulse fraction.
The vertical bars show the 1$\sigma$ level error range.
Fluxes obtained on the same day are connected by solid lines.
For the Subaru Telescope data, we used the results of \citet{Morii+_2005}
and \citet{Tanaka+_2008}.
To convert the magnitudes to fluxes,
we used the zero fluxes of the Johnson-Cousins photometric system
for the $I$$R$$V$$B$ bands of the KeckI(II)/LRIS, KeckII/ESI and UH88/Optic data,
the Ellis system for the $R_E$ band of the KeckII/ESI data,
the SDSS system for the $i^\prime$$g^\prime$ bands of the UHT/ULTRACAM data
\citep{Fukugita_Shimasaku_Ichikawa_1995}
and
the MKO-NIR system for the $K$$K^\prime$$Ks$ bands of the
KeckI/NIRC, CFHT/AOB, Subaru/IRCS and Gemini/NIRI data
\citep{Tokunaga_Vacca_2005}.
We corrected the observed flux by the interstellar
extinction of $A_V = 3.5$ \citep{Durant_van_Kerkwijk_2006a}.
Extinction other than that of the $V$ band was calculated by the relative extinction
listed in \citet{Schlegel_Finkbeiner_Davis_1998} and
the extinction law given in \citet{Chiar_Tielens_2006}.
The systems or bands not listed in \citet{Schlegel_Finkbeiner_Davis_1998}
were corrected using values of the Landolt and UKIRT systems
for optical and near-infrared bands, respectively.
    }\label{fig: 4u_nufnu_av3.5+pulse.ps}
\end{figure}

To discuss the emission mechanism of the AXP, 
we generated the spectral energy distribution ($\nu F_\nu$) of
4U 0142+61 for the phase-averaged fluxes and
pulsed fluxes (figure \ref{fig: 4u_nufnu_av3.5+pulse.ps}),
using all the published data (see caption of figure \ref{fig: 4u_nufnu_av3.5+pulse.ps})
and the $K^\prime$ band fluxes obtained in this work.

For the Subaru observation on Sep. 2003 \citep{Morii+_2005},
we analysed the data in our own way
and obtained different results from that shown in \citet{Durant_van_Kerkwijk_2006b}.
We will present the details of the photometry
in our companion paper \citep{Tanaka+_2008}.
In the phase-averaged spectrum, a peak at the $H$ band is clearly seen.
As discussed in \citet{Hulleman_van_Kerkwijk_Kulkarni_2004},
this feature may be a proton cyclotron feature.
Although other cyclotron features in other bands
are expected, they are not yet clear.
Further simultaneous multi-band observations are necessary,
because the spectrum is variable.

To understand the emission mechanism of the pulse component,
we calculated the slope of the spectral shape.
Combined with the pulsed flux of
the $K^\prime$ band (this work; $PF_{\rm rms}$)
and that of the $i^\prime$ band \citep{Dhillon+_2005},
the slope of the pulsed component ($F_\nu \propto \nu^\alpha$) was
found to be constrained to $\alpha > -0.87$ (90\% C.L.) for the interstellar
extinction of $A_V = 3.5$.

%
Because the constraint on the slope of the pulsed flux is not so stringent,
any emission mechanisms are still allowed for the pulse component.
For example, the Rayleigh-Jeans tail of the blackbody is $\alpha = +2$ and
synchrotron self-absorption is $\alpha = +2.5$.
In the quantum electrodynamics model of \citet{Heyl_Hernquist_2005},
$\alpha = 0$ for the optical-infrared region.
In the disk-star dynamo gap model of \citet{Ertan_Cheng_2004}, $\alpha = -0.5$;
while in the magnetar model by the same authors, $\alpha$ is slightly smaller than $-0.5$.

In the curvature-drift-induced maser mechanism proposed by \citet{Lu_Zhang_2004},
the maser frequency of 4U 0142+61 is about $\nu_M = 1.39 \times 10^{14}$ Hz,
which is near the $K$ band. 
Then, the maser amplification ratio equating to the pulse fraction in their model
is maximum near the $K$ band.
Since the upper limit for the pulse fraction of the $K^\prime$ band
is similar to the pulse fraction of the $i^\prime$ band,
this model is a possible mechanism.
The $H$ band peak in the phase-averaged spectrum may be the maser frequency peak of
this emission mechanism. If so, then a larger pulse fraction at the $H$ band is expected.

Viewed differently,
the dust disk model is also a possible mechanism
\citep{Wang_Chakrabarty_Kaplan_2006}, because
the fraction of un-pulsed flux is dominant in the $K^\prime$ band.


In the near future, we can make use of the adaptive optics (AO)
technique for the sensitive detection of the AXP. The Subaru Telescope
plans to install an upgraded 188-element AO system with a laser guide
star \citep{Hayano+_2003}.
Using this technique, we will be able to constrain the pulse fraction
to $PF < 3\%$ and constrain the spectral slope of the pulsed component
to $\alpha > 1$ for a feasible observation time.
Furthermore, the $K$ band flux varies about 40\% 
over the time scale of years and hence
it is possible that the $K$ band pulsation would grow above the detection limit.
Further monitoring of the infrared pulsation should prove interesting.


\bigskip

M.~M. is supported by the Post-doctoral Researchers Program of Rikkyo University and
a Grant-in-Aid for Young Scientists(B)(18740120).


\end{document}